\documentclass[prl,twocolumn,amsmath,amssymb]{revtex4} 

\usepackage{graphicx}
\usepackage{subfigure} 
\usepackage{dcolumn}
\usepackage{array} 
\usepackage{bm}
\usepackage[hang]{footmisc} 
\def\ub{\underline} 

\usepackage[usenames,dvipsnames]{color}

\begin{document} 
 
\title{\bf{Orthorhombic $ABC$ semiconductors as antiferroelectrics} \\[11pt] } 
\author{Joseph W. Bennett$^{*}$, Kevin F. Garrity$^{*}$, Karin M. Rabe and David Vanderbilt} 
\affiliation{Department of Physics and Astronomy\\ 
Rutgers University, Piscataway, NJ 08854} 
\affiliation{$^{*}$These authors contributed equally to the present work.} 
 
\date{\today} 
\begin{abstract} 
We use a first-principles rational-design approach to identify a 
previously-unrecognized class of antiferroelectric materials in the 
$Pnma$ MgSrSi structure type. The MgSrSi structure type can be 
described in terms of antipolar distortions of the nonpolar 
$P6_{3}/mmc$ ZrBeSi structure type, and we find many members of this 
structure type are close in energy to the related polar $P6_{3}mc$ 
LiGaGe structure type, which includes many members we predict to be 
ferroelectric. We highlight known $ABC$ combinations in which this 
energy difference is comparable to the antiferroelectric-ferroelectric 
switching barrier of PbZrO$_{3}$. We calculate structural parameters 
and relative energies for all three structure types, both for reported 
and as-yet hypothetical representatives of this class. Our results 
provide guidance for the experimental realization and further 
investigation of high-performance materials suitable for practical 
applications. 
\end{abstract} 
 
 
\maketitle 
 

There has been great progress in the use of first-principles methods
in the design and discovery of new functional materials, most notably
for the rapid evaluation and screening of a large number of known and
as-yet hypothetical compounds in a target family already known to
contain compounds exhibiting the desired functional
behavior\cite{Jain11p2295, Armiento11p014103, Zhang12p1425,
  Roy12p037602}. A greater challenge is to develop ways to identify
new functional materials in families in which the desired behavior has
been previously unrecognized.  In recent work, we used a combined
crystallographic database/first-principles approach
\cite{Bennett12p167602} to identify semiconducting members of the
family of compounds in the LiGaGe structure type as a previously
unrecognized class of ferroelectrics, characterized by spontaneous
polarizations and barriers to polarization switching comparable to the
much-studied ferroelectric $AB$O$_{3}$ perovskite
oxides~\cite{Bennett12p167602, Bennett12p21}.  Here,
we apply this approach to identify a previously unrecognized class of
antiferroelectrics, opening the way to increased recognition and
application of antiferroelectrics as functional materials.

An antiferroelectric \cite{Lines77,Rabe12AFE} is like a ferroelectric
in that its structure is obtained through distortion of a nonpolar
high-symmetry reference phase; for ferroelectrics, the distortion is
polar, while for antiferroelectrics it is nonpolar.  However, not all
nonpolar phases thus obtained are antiferroelectric: in addition,
there must be an alternative low-energy ferroelectric phase obtained
by a polar distortion of the same high-symmetry reference structure,
and an applied electric field must induce a first-order transition
from the antiferroelectric phase to this ferroelectric phase,
producing a characteristic P-E double-hysteresis loop
\cite{Shirane51p476,Kittel51p729}.  This behavior in applied electric
field is the origin of the functional properties of
antiferroelectrics.  For systems in which there is a difference in
lattice parameters between the AFE and FE phases, the
electric-field-induced transition produces a large nonlinear strain
response, useful for transducer applications~\cite{Yu01p333}.  The
entropy change between the two phases similarly can produce a high
effective electrocaloric response, useful for solid-state
cooling~\cite{Mischenko06p1270}.  Finally, the shape of the $P$-$E$
hysteresis loop allows storage of electrical energy of sufficient
magnitude to attract interest for energy storage applications
~\cite{Jaffe61p1264,Tan11p4091}.
 
Most of the research on antiferroelectric oxides has focused on two
classes of materials, both with rather complex structures. The $Pbam$
structure of the prototypical perovskite antiferroelectric
PbZrO$_3$\cite{Shirane51p476} is obtained through a cell-quadrupling
antipolar Pb displacement mode combined with four additional modes
that further double the unit cell~\cite{Fujishita82p3583}; the $Pbcm$
structure of the antiferroelectric phases of NaNbO$_3$ and AgNbO$_3$
are similarly complex~\cite{Mishra07p024110}. Chemical substitution
into the endpoint compounds allows tuning of the critical temperature,
the critical field, the electric-field induced strain and
polarization, and other functional properties~\cite{Tan11p4091,
  Jankowska-Sumara07p1887}.  Chemical substitution is also seen to
induce transitions to a distinct tetragonal antiferroelectric phase,
for example in
Pb$_{1-x}$Sr$_{x}$(Zr$_{1-y}$Ti$_{y}$)O$_{3}$~\cite{Yu01p333}, and to
related ferroelectric phases, for example in PbZr$_{1-x}$Ti$_x$O$_3$,
also obtained as distortions of the ideal perovskite structure.

Evidence is beginning to emerge that antiferroelectricity in inorganic
materials extends far beyond the systems which have
been the focus of the literature to date. In particular, recent
observations of orthorhombic $Pnma$ Sm-doped
BiFeO$_3$~\cite{Kan10p1108} and thin-film
BiCrO$_3$~\cite{Kim06p162904} show the double hysteresis loops
characteristic of antiferroelectricity.  This suggests that many additional
nonpolar phases should be considered as candidates for
previously-unrecognized antiferroelectricity.  Our strategy to search
for new antiferroelectric materials is to identify a class of
materials with a structure type that is obtained from an antipolar
distortion of a high-symmetry reference structure which is also
related through a polar distortion to a structure type of known
ferroelectrics.  By screening a large number of both reported and
hypothetical compounds, we can find compositions that are insulating
and locally stable both in the antiferroelectric structure and in the
related ferroelectric structure; of these, the compounds for which the
antiferroelectric structure is slightly lower in energy than the
ferroelectric structure are promising candidates for new
antiferroelectrics for targeted experimental investigation.  Through
this search, we expect to find new AFE structure types which are
significantly simpler than the $Pbam$ structure, even perhaps including
realizations of the two-sublattice Kittel model~\cite{Kittel51p729}
and $q$=0 antiferroelectrics~\cite{Blinc74soft}.  For technological
applications, new materials offer the possibility of better
performance via reduced hysteresis, larger changes at the
electric-field-induced transition, operation at higher or lower
temperatures, better integration with other materials based on
structural or chemical compatibility, elimination of toxic elements,
and/or a band gap suitable for photoactive applications.
 
\begin{figure} 
  \centering 
  \includegraphics[width=3.5in]{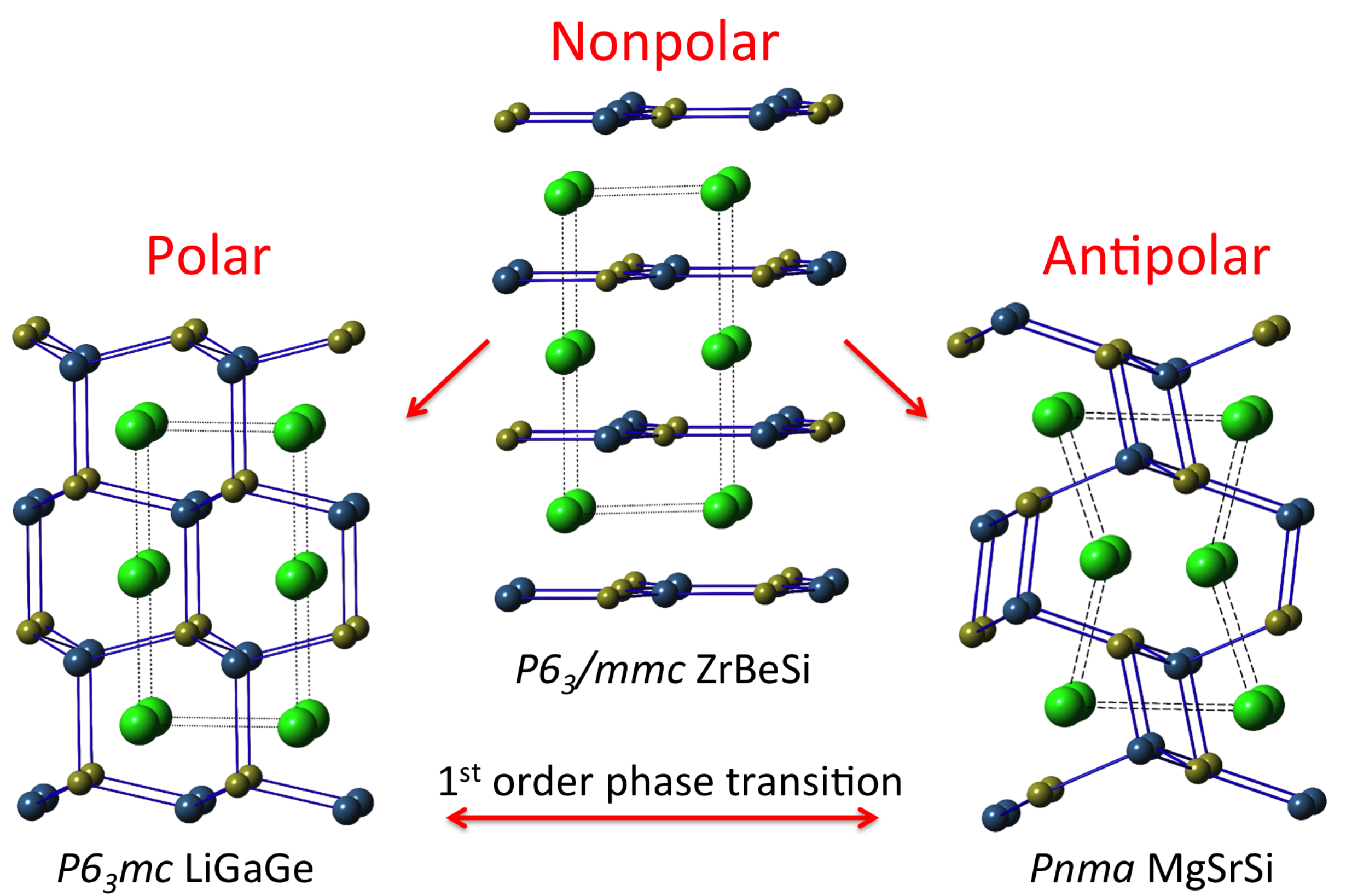} 
  \caption{Left: planar buckling distortions of the polar
  $P6_{3}mc$ LiGaGe structure relative to the $P6_{3}/mmc$
  high symmetry ZrBeSi structure type (center), as described in
  Ref.~\cite{Bennett12p167602}. Right: antipolar distortions of
  the $Pnma$ MgSrSi structure type relative to the high symmetry
  ZrBeSi structure type (center), showing antipolar buckling
  of the planes formed by atoms $BC$ at Wyckoff positions $2b$
  (dark blue) and $2b^\prime$ (gold) and antipolar displacements
  of the stuffing atoms $A$ at $2a$ (green).}
  \label{fig:energetics} 
  \end{figure}

$ABC$ compounds in the $Pnma$ MgSrSi structure
type~\cite{Eisenmann72p241} (previously referred to as the 
anti-PbCl$_{2}$ structure type~\cite{Shoemaker65p900}) are a
promising target class. The MgSrSi structure, shown in Fig.~\ref{fig:energetics}, is
obtained by a nonpolar distortion of the high-symmetry $P6_{3}/mmc$
ZrBeSi structure.  The distortion can be decomposed into two modes: an
$M_{2}^{-}$ mode (antipolar displacements of $ABC$ along $c$, and
antipolar displacements of $A$ along $b$) that breaks the $P6_{3}/mmc$
symmetry to $Pnma$, followed by a $\Gamma_{5}^{+}$ mode (shifting of
$BC$ layers along $b$) that does not break any additional
symmetry. A polar distortion specified by a $\Gamma_{2}^{-}$ mode
relates the high-symmetry $P6_{3}/mmc$ structure to the LiGaGe-type
$P6_{3}mc$ ferroelectric compounds identified
previously~\cite{Bennett12p167602}; in addition to being
insulating, these compounds can have spontaneous polarization
comparable to that of BaTiO$_3$.  However, little
has been reported about the band structure of
MgSrSi-type compounds or about their response to applied electric
fields~\cite{Liu06p830,
Katsura12preprint}.

In this paper, we use first-principles methods to establish a new
class of antiferroelectrics in the MgSrSi structure type and to
identify promising candidate materials for further
investigation. Specifically, we compute the structural parameters,
nonpolar distortions, band gaps, and AFE-FE energy differences for a
search set comprising 37 reported and 33 as-yet-hypothetical $ABC$
compounds in the MgSrSi structure type. We identify 11 combinations
for which both the FE and AFE phases are insulating and have an energy
difference below 0.2~eV. For all insulating combinations studied, we
find that the band gaps are in the semiconducting range; the lower
band gaps could be useful for photoactive
applications~\cite{Bennett08p17409, Bennett10p184106,
  Gou11p205115}. These candidate antiferroelectrics offer promise for
experimental investigation and for the future development of new
high-performance materials for practical applications.
 
First principles computations were performed with the ABINIT
package~\cite{Gonze09p2582}.  The local density approximation (LDA)
and a 6$\times$6$\times$6 Monkhorst-Pack sampling of the Brillouin
zone~\cite{Monkhorst76p5188} were used for all calculations, except
for the Berry phase polarization~\cite{KingSmith93p1651, Resta94p899}
calculations, for which an 8$\times$8$\times$8 grid was used. All
atoms were represented by norm-conserving,
optimized~\cite{Rappe90p1227}, designed nonlocal~\cite{Ramer99p12471}
pseudopotentials, generated with the OPIUM code~\cite{Opium}. All
calculations were performed with a plane wave cutoff of 50~Ry.  In
addition, the QUANTUM ESPRESSO~\cite{Giannozzi09p395502} package was
used to perform nudged elastic band
calculations~\cite{neb,nebci}. WANNIER90~\cite{wannier90} was used to
generate maximally localized Wannier functions (MLWF)~\cite{mlwf}.
 
 Our search set of candidate MgSrSi-type antiferroelectrics consists of
 $ABC$ combinations drawn from the {\it Inorganic Crystal Structural
   Database} (ICSD)~\cite{Belsky02p364} and two recent high-throughput
 searches.  The ICSD includes 37 previously synthesized $ABC$
 compounds in the MgSrSi structure type that do not contain an
 $f$-block element and which have a total of 8 or 18 valence
 electrons, which promotes band gap formation.  In addition, we
 included 13 as-yet unsynthesized compounds from the high-throughput study of
 Zhang {\em et al.}~\cite{Zhang12p1425} that are predicted to have the
 desired $Pnma$ structure.  Finally, we included 20 compounds
 from the high-throughput study of Bennett {\em et
   al.}~\cite{Bennett12p167602} that are predicted to be insulating
 and at least locally stable in the polar LiGaGe
 structure type, bringing our total search
 set to 70 compounds.  We can classify these combinations into the
 following groups: I-I-VI (NaKSe), I-II-V (LiCaBi), I-XII-V (NaCdAs),
 XI-II-V (CuMgP), II-II-IV (SrCaGe), III-X-V (ScNiP), III-XI-IV
 (ScCuSi), and IV-X-IV (ZrPtSi).

 For each $ABC$ combination in our search set, we optimize the
 structural parameters for each of the three structural variants
 $\ub{A}BC$, $A\ub{B}C$, and $AB\ub{C}$, where the underscore
 indicates the stuffing atom (Fig.~\ref{fig:energetics}), for each of
 our three structure types: MgSrSi ($Pnma$), LiGaGe ($P6_{3}mc$), and
 ZrBeSi ($P6_{3}/mmc$). The computed structural parameters
 (Supplemental Table 1) generally show good agreement with
 experimental values, with the underestimate of lattice constants
 characteristic of LDA calculations, about 1-3~$\%$ for $a$ and as
 large as 3-4~$\%$ for $c$\footnote{For select combinations in which
   the error in lattice constant is larger, on the order of 3-4\%, we
   perform all-electron calculations as implemented in
   Wien2K~\cite{Schwarz03p259}, and obtain lattice constants similar
   to those predicted (Supplemental Materials) using the Bennett-Rappe
   library of pseudopotentials (design parameters described in
   Ref.~\cite{Bennett12p14}).}. While most of our calculations agree
 with previous experimental and theoretical determinations of the
 ground state structure, there are a few minor discrepancies.  We find
 that NaCdSb prefers the $P6_{3}mc$ structure by 34~meV/f.u.\ over the
 experimentally reported $Pnma$ structure. In addition, we find that
 KSrBi and RbBaAs prefer the high-symmetry $P6_{3}/mmc$ structure, in
 contrast to the results of Zhang {\rm et al.}\cite{Zhang12p1425},
 which we attribute to minor methodological differences~\footnote{We
   believe that these few discrepancies, as well as those combinations
   in which only one refinement was present in the ICSD, warrant
   further experimental investigations.}.
 
 Our most promising candidate antiferroelectrics, shown in
 Table~\ref{table:insulators}, fulfill the following criteria: a) the
 ground state is the antipolar $Pnma$ MgSrSi structure type, b) the
 polar $P6_{3}mc$ LiGaGe structure type is metastable, and c) both phases
 are insulating. In Table~\ref{table:insulators} we present the energy
 differences between the $Pnma$ and $P6_{3}mc$ structures ($\Delta
 E$), the energy difference between the $P6_{3}mc$ and $P6_{3}/mmc$
 ($\Delta E_{\rm SW}$), the polarization of the LiGaGe structure, the
 mode decomposition of the antipolar distortions, and difference in
 volume between the $P6_{3}mc$ and $Pnma$ structures. Of this set,
 seven combinations are known materials, highlighted in boldface in
 Table~\ref{table:insulators}.  The calculated band gaps of these
 materials cover the entire semiconducting range, from 0.04~eV
 (BaCaSi) to 2.91~eV (NaLiTe), with changes of 0.01 to 0.59~eV upon
 switching (see Supplemental Tables).

\begin{table} 
\begin{center} 
\begin{ruledtabular} 
\begin{tabular}{lccccccc} 
$ABC$ & $\Delta E_{\rm SW}$ & $P$ & $\Gamma_{5}^{+}$ &$M_{2}^{-}$& $\Delta E$ & $\Delta V$& $\Delta V/V$\\ 
      & (meV)              & (C/m$^{2}$)&       &           & (meV)       &  (\AA$^{3}$) & (\%)\\  
\hline
\ub{Li}BeP       & 119 & 0.85 & 0.26 & 1.26 & 18  & 1.23 & 3.7\\ 
\ub{Mg}LiP       & 20  & 0.38 & 0.41 & 1.28 & 230 & 0.49 & 1.1\\ 
\ub{Mg}LiAs      & 30  & 0.39 & 0.42 & 1.35 & 207 & 0.12 & 0.2\\ 
{\bf\ub{Ca}LiSb} & 7   & 0.18 & 0.19 & 1.17 & 79 & 0.09 & 0.1\\ 
{\bf\ub{Ca}LiBi} & 7   & 0.19 & 0.20 & 1.16 & 80  & $-$0.10 & $-$0.1\\ 
\ub{Na}MgP       & 102 & 0.49 & 0.31 & 1.43 & 275 & 1.88 & 3.2\\ 
\ub{Na}MgAs      & 114 & 0.48 & 0.34 & 1.44 & 232 & 1.65 & 2.5\\ 
\ub{Na}MgSb      & 146 & 0.43 & 0.40 & 1.51 & 154 & 0.93 & 1.2\\ 
\ub{Na}MgBi      & 127 & 0.42 & 0.44 & 1.51 & 143 & 1.14 & 1.4\\ 
\ub{K}MgSb       & 41  & 0.40 & 0.33 & 1.77 & 254 & 2.17 & 2.6\\ 
\ub{K}MgBi       & 73  & 0.31 & 0.34 & 1.78 & 227 & 1.65 & 1.8\\ 
\ub{Na}ZnSb      & 81  & 0.49 & 0.18 & 1.48 & 42  & 2.68 & 4.1\\ 
{\bf\ub{Na}LiTe} & 23  & 0.20 & 0.30 & 1.41 & 114 & 0.96 & 1.4\\ 
{\bf\ub{K}NaS}   & 12  & 0.17 & 0.32 & 1.36 & 149 & 1.60 & 2.3\\ 
{\bf\ub{K}NaSe}  & 13  & 0.15 & 0.33 & 1.38 & 131 & 2.09 & 2.7\\ 
{\bf\ub{K}NaTe}  & 14  & 0.13 & 0.38 & 1.42 & 96  & 2.81 & 2.9\\ 
\ub{Na}AgSe      & 51  & 0.65 & 0.62 & 1.63 & 134 & 0.11 & 0.2\\ 
{\bf\ub{Ba}CaSi} & 11  & 0.34 & 0.35 & 1.62 & 310 & 0.37 & 0.4\\ 
\end{tabular} 
\end{ruledtabular} 
\caption{Energy difference
  between $P6_{3}mc$ and $P6_{3}/mmc$ phases ($\Delta E_{\rm SW}$),
  polarization $P$, amplitude of the microscopic distortions
  from ZrBeSi to MgSrSi structure types ($\Gamma_{5}^{+}$ and
  $M_{2}^{-}$), energy difference between AFE and FE states
  ($\Delta E$), and change in volume ($\Delta V$), also
  given as a percentage, for the eighteen compounds in the search set 
  that are insulating in both the
  antipolar and polar states. Combinations reported as $Pnma$ in the ICSD
  are shown in boldface. Energies are reported in meV per formula unit.  Band
  gaps are reported in the supplemental material.}
\label{table:insulators} 
\end{center} 
\end{table}

\begin{table} 
\begin{center} 
\begin{ruledtabular} 
\begin{tabular}{lccccccc} 
$ABC$ & $\Delta E_{\rm SW}$ & $P$ & $\Gamma_{5}^{+}$ &$M_{2}^{-}$& $\Delta E$ & $\Delta V$& $\Delta V/V$\\ 
      & (meV)              & (C/m$^{2}$)&       &           & (meV)       &  (\AA$^{3}$) & (\%)\\   
\hline
\ub{Mg}LiSb      & 50  & ---  & 0.48 & 1.42 & 142 & 1.32 & 2.2\\ 
\ub{Mg}LiBi      & 41  & ---  & 0.52 & 1.45 & 126 & 1.52 & 2.4\\ 
{\bf\ub{Mg}CuP}  & 1   & 0.36 & 0.42 & 1.43 & 174 & $-$0.58 & $-$1.3\\ 
\ub{Na}ZnBi      & 328 & ---  & 0.18 & 1.28 &  18 & 2.50 & 3.6\\ 
{\bf\ub{Na}CdAs} & 199 & 0.40 & 0.31 & 1.44 &  50 & 1.54 & 2.4\\ 
{\bf\ub{Sr}CaGe} & 2   & 0.04 & 0.27 & 1.36 & 186 & 3.72 & 4.6\\ 
{\bf\ub{Ba}CaGe} & 11  & 0.26 & 0.37 & 1.59 & 289 & 1.04 & 1.1\\ 
\end{tabular} 
\end{ruledtabular} 
\caption{Same as Table 1, but for compounds in the search set for which
  either the $Pnma$ state {\it or} the $P6_3mc$ state is metallic in our DFT
  calculation.  A dash
  in the polarization ($P$) column indicates that
  the $P6_{3}mc$ state is metallic.}
\label{table:metals} 
\end{center} 
\end{table} 

We find eleven insulating compounds with energy differences between
the polar and antipolar phases below 0.2~eV/f.u. (see
Table~\ref{table:insulators}). Of this set, six combinations, namely
LiCaBi, LiCaSb, KNaTe, NaLiTe, KNaS, and KNaSe, are known
materials. Despite the small energy differences, many of our candidate
materials still require a large critical field to stabilize the polar
phase. However, the structural versatility of $ABC$ intermetallic
compounds should make it experimentally feasible to reduce the
critical field of many of these materials via formation of solid
solutions of chemically similar compounds with different
structures. For instance, we find that LiBeP has an antipolar $Pnma$
ground state, while LiBeAs has a polar $P6_{3}mc$ ground state.  A
solid solution of these two materials should result in a material in
which it is possible to chemically tune towards the first-order
transition between preferred structures, thus reducing the critical
field.

 \begin{figure} 
 \centering 
 \includegraphics[width=3.5in]{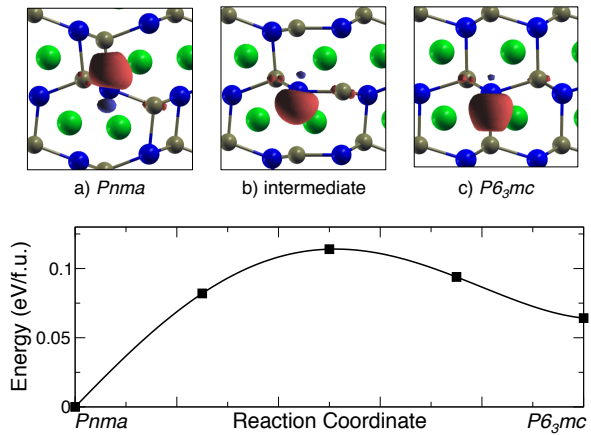} 
 \caption{Minimum-energy path for switching between the non-polar
 $Pnma$ and polar $P6_{3}mc$ phases of NaCdAs, as calculated by
 the NEB method. (a-c):
 isosurfaces of an inter-planar bonding
 state (MLWF), which reverses orientation during the switching
 process. Lower panel: graph of energy vs reaction coordinate,
 showing that our
 proposed reaction pathway from $Pnma$ to $P6_{3}mc$ has a
 barrier of 0.11~eV. This is in comparison to a path
 that would have the high-symmetry $P6_{3}/mmc$ structure as our
 intermediate, which would be 0.19~eV.
}
 \label{fig:neb_wan}
 \end{figure}

In addition to our most promising candidates, in
Table~\ref{table:metals} we present data for seven compounds that have
a $Pnma$ ground state as well as a stable polar distortion, but for
which only one of the two phases is insulating.  This caveat would
normally disqualify these compounds as candidate antiferroelectrics,
but it is well known that DFT-LDA tends to significantly underestimate
band gaps, and can even predict small-gap semiconductors (e.g., Ge) to
be metallic~\cite{Bachelet85p879}.  It is possible that these seven
compounds could display antiferroelectric behavior, and therefore
merit experimental investigation, especially since four are known
materials (highlighted in boldface in Table~\ref{table:metals}).

In order to investigate the switching path between the antipolar and
polar states, we use nudged elastic band calculations (with the volume
fixed to the $Pnma$ cell) to identify the lowest-energy path between
the two phases, as shown in Fig.~\ref{fig:neb_wan}.  While we expect
that the actual experimental switching path will depend on complex kinetic
processes like domain nucleation and domain-wall motion, our
calculations of homogeneous switching paths provide an estimate of the
energy scales involved.  For our example of NaCdAs in
Fig.~\ref{fig:neb_wan}, we find that the low-energy switching path
from $Pnma$ to $P6_{3}mc$ corresponds primarily to half of the Cd
atoms moving from below the As layer to above it, while the Na atoms
rearrange to avoid the Cd. For NaCdAs, this first-order transition path
reduces the barrier to switching by more than 50\% as compared to
switching through the high-symmetry $P6_{3}/mmc$ structure, and makes
it more likely that the material will return to the ground state when
the external field is switched off.

We use maximally localized Wannier functions to analyze differences in
bonding between the $P6_{3}mc$ and $Pnma$ structures and to study how
the bonding changes when the material is switched.  While we refer to
the previously synthesized NaCdAs as our example, we find that our
description of bonding is similar in other combinations. The bonding
orbitals of the $P6_{3}mc$ structure are $sp^3$-like orbitals centered
near As, the most electronegative element, with all bonds oriented
towards the four neighboring Cd atoms (see Fig.~\ref{fig:neb_wan}).  The
bonds of the $Pnma$ structure are also $sp^3$-like and centered on As;
however, a transition between the two structures requires that half of
the $sp^3$ bonds reorient.  This proceeds by breaking an
inter-layer Cd-As bond, with this orbital forming a $p_z$-like
non-bonding state in the transition structure, and then reforming the
bond pointing in the other direction (see the orbital in
Fig.~\ref{fig:neb_wan}).  In addition, the in-plane $sp^3$ bonds
reorient by passing through an $sp^2$-like intermediate state.  The
similar bonding in the $P6_{3}mc$ and $Pnma$ structures is consistent
with their competitive energies. However, we note that in the $Pnma$
structure, only half of the $sp^3$ bonds are oriented directly between
Cd and As, which accounts for the significant changes in band gap
between the two structure types.

Our proposed antiferroelectrics have many properties that could make
them better for applications than existing materials.  For instance,
most of our candidates have a change in volume of 1-4\,\% upon
switching (see Table \ref{table:insulators}), which is much larger
than the change observed in most piezoelectrics (0.1-0.2\,\%) and
current antiferroelectrics ($\le$ 0.9\,\%)~\cite{Yu01p333}. Very large
volume changes could make these materials ideal for a variety of
transducer applications in which harnessing a large non-linear
response is necessary. Another application would be to use these
antiferroelectrics as a substrate which could apply a reversible
strain to thin films grown on top. We also note that in contrast to
perovskite oxides, the stability of the polar and antipolar
distortions in these materials is relatively insensitive to
strain~\cite{Bennett12p167602}, these materials have no competing
distortion patterns, and they are strongly anisotropic, a combination
of favorable properties which could make antiferroelectric $ABC$
useful in applications that require a robust and reversible response
to external fields, such as high-energy storage capacitance and
electrocaloric refrigeration.

In conclusion, we have used first-principles methods to establish a
new class of antiferroelectrics in the MgSrSi structure type and to
identify promising candidate materials for further investigation.
Through targeted synthesis, MgSrSi-type compounds could potentially be
developed as a valuable class of functional antiferroelectric
materials. This is a specific application of a larger-scale strategy
to identify new functional materials by targeting insulating compounds
not previously recognized as functional materials and tuning the
composition and other control parameters, such as epitaxial strain,
and/or modifying the structure by intercalation of atoms.  The identification of
antiferroelectricity in classes of materials in which it was
previously unrecognized offers the possibility of optimizing
properties and combining polarization with other functional
properties, including magnetism, to produce multifunctional behavior
of fundamental scientific interest and for groundbreaking
technological applications.
 
\vspace{0.3cm} 
\noindent{\bf Acknowledgments} 
\vspace{0.3cm} 
 
This work was supported in part by ONR Grants N00014-09-1-0302 and 
N00014-05-1-0054. Calculations were carried out at the Center for 
Piezoelectrics by Design. We thank P. K. Davies, D. R. Hamann and 
R. Seshadri for useful discussions. K. M. R. thanks R. Seshadri for 
hospitality at UCSB and the Aspen Center for Physics (NSF Grant 
1066293) where part of this work was carried out. 
  
\bibliography{thebibliography}

\end{document}


\title{\bf{Supplemental material for Orthorhombic $ABC$ semiconductors as antiferroelectrics} \\[11pt] }
\author{Joseph W. Bennett, Kevin F. Garrity, Karin M. Rabe and David Vanderbilt}
\affiliation{Department of Physics and Astronomy\\
Rutgers University, Piscataway, NJ 08854 }

\date{\today}
\maketitle

\begin{table}
\centering
\begin{tabular}{lcccccccccc}
$ABC$&$a$&$b$&$c$&$x_{\rm A}$&$z_{\rm A}$&$x_{\rm B}$&$z_{\rm B}$&$x_{\rm C}$&$z_{\rm C}$&$E_{\rm gap}$\\
     &(\AA)&(\AA)&(\AA)&&&&&&&(eV)\\
\ub{Na}LiSe &7.09 (7.17) &4.19 (4.25) &7.54 (7.72) &0.011 (0.014) &0.684 (0.684) &0.143 (0.152) &0.072 (0.072) &0.282 (0.277) &0.394 (0.394) &2.84\\
\ub{Na}LiTe &7.61 (7.74) &4.52 (4.62) &8.16 (8.41) &0.012 (0.013) &0.686 (0.695) &0.143 (0.157) &0.072 (0.075) &0.278 (0.249) &0.393 (0.393) &2.91\\
\ub{K}NaS   &7.59 (7.70) &4.51 (4.60) &8.31 (8.29) &0.017 (0.013) &0.682 (0.685) &0.146 (0.147) &0.074 (0.076) &0.276 (0.279) &0.391 (0.396) &2.59 \\
\ub{K}NaSe  &7.88 (8.06) &4.70 (4.82) &8.65 (8.65) &0.017 (0.013) &0.683 (0.685) &0.147 (0.147) &0.073 (0.075) &0.274 (0.279) &0.391 (0.396) &2.21 \\
\ub{K}NaTe  &8.35 (8.52) &5.00 (5.13) &9.26 (9.26) &0.017 (0.013) &0.686 (0.685) &0.147 (0.147) &0.072 (0.075) &0.272 (0.279) &0.390 (0.393) &2.41\\
\ub{Ca}LiSb &7.46 (7.52) &4.53 (4.67) &8.07 (8.50) &0.007 (0.007) &0.705 (0.702) &0.154 (0.188) &0.074 (0.074) &0.270 (0.268) &0.403 (0.401) &0.58 \\
\ub{Ca}LiBi &7.57 (7.77) &4.59 (4.75) &8.21 (8.44) &0.008 (0.006) &0.705 (0.704) &0.154 (0.113) &0.073 (0.083) &0.270 (0.269) &0.402 (0.401) &0.48 \\
\ub{Sr}LiSb &7.86 (8.04) &4.71 (4.82) &8.24 (8.46) &0.002 (0.003) &0.709 (0.709) &0.160 (0.161) &0.076 (0.076) &0.278 (0.275) &0.407 (0.407) &0.62 \\
\ub{Sr}LiBi &7.95 (8.29) &4.78 (4.93) &8.37 (8.76) &0.002 (0.004) &0.709 (0.707) &0.160 (0.176) &0.076 (0.094) &0.277 (0.275) &0.407 (0.407) &0.54 \\
\ub{Mg}CuP  &6.48 (6.53) &3.81 (3.84) &7.07 (7.17) &0.028 (0.032) &0.680 (0.679) &0.128 (0.128) &0.061 (0.063) &0.258 (0.252) &0.378 (0.379) &$m$  \\
\ub{Na}CdAs &7.49 (7.57) &4.42 (4.47) &8.03 (8.04) &0.017 (0.016) &0.675 (0.673) &0.150 (0.150) &0.074 (0.075) &0.278 (0.279) &0.390 (0.391) &$m$  \\
\ub{Na}CdSb &7.90 (7.95) &4.67 (4.71) &8.43 (8.42) &0.014 (0.011) &0.680 (0.674) &0.150 (0.150) &0.076 (0.075) &0.277 (0.278) &0.394 (0.394) &$m$ \\
\ub{Sc}NiP  &6.26 (6.33) &3.68 (3.73) &7.00 (7.08) &0.021 (0.022) &0.684 (0.685) &0.139 (0.142) &0.060 (0.062) &0.266 (0.265) &0.381 (0.381) &$m$ \\
\ub{Sc}PtP  &6.47 (6.44) &3.78 (4.29) &7.30 (7.55) &0.029 (0.023) &0.679 (0.687) &0.148 (0.157) &0.061 (0.065) &0.250 (0.264) &0.375 (0.387) &0.04 \\
\ub{Sc}CuSi &6.46 (6.57) &3.93 (3.98) &7.12 (7.22) &0.007 (0.009) &0.697 (0.695) &0.160 (0.158) &0.069 (0.066) &0.275 (0.270) &0.393 (0.390) &$m$ \\
\ub{Ca}MgSi &7.37 (7.49) &4.37 (4.44) &8.29 (8.33) &0.021 (0.019) &0.680 (0.679) &0.145 (0.143) &0.063 (0.063) &0.265 (0.270) &0.383 (0.385) &$m$ \\
\ub{Ca}MgGe &7.39 (7.62) &4.40 (4.42) &8.34 (8.34) &0.021 (0.019) &0.680 (0.680) &0.146 (0.143) &0.064 (0.060) &0.265 (0.270) &0.384 (0.383) &0.23 \\
\ub{Ca}MgSn &7.75 (7.86) &4.63 (4.66) &8.76 (8.74) &0.023 (0.018) &0.682 (0.683) &0.144 (0.148) &0.062 (0.061) &0.261 (0.268) &0.382 (0.386) &$m$ \\
\ub{Sr}MgSi &7.71 (7.78) &4.54 (4.56) &8.43 (8.49) &0.016 (0.015) &0.684 (0.683) &0.146 (0.140) &0.065 (0.057) &0.277 (0.276) &0.389 (0.390) &$m$ \\
\ub{Sr}MgGe &7.75 (7.80) &4.57 (4.56) &8.46 (8.55) &0.015 (0.014) &0.684 (0.684) &0.147 (0.142) &0.066 (0.064) &0.279 (0.278) &0.389 (0.390) &$m$ \\
\ub{Sr}MgSn &8.08 (8.18) &4.80 (4.92) &8.84 (8.75) &0.014 (0.011) &0.688 (0.689) &0.148 (0.144) &0.067 (0.066) &0.273 (0.278) &0.391 (0.392) &$m$ \\
\ub{Sr}CaSi &7.87 (8.11) &4.82 (4.94) &8.91 (9.17) &0.021 (0.021) &0.681 (0.679) &0.149 (0.149) &0.075 (0.074) &0.267 (0.266) &0.395 (0.394) &0.20  \\
\ub{Sr}CaGe &7.89 (8.12) &4.84 (4.95) &8.92 (9.18) &0.020 (0.021) &0.682 (0.679) &0.150 (0.148) &0.076 (0.073) &0.266 (0.267) &0.396 (0.393) &0.23 \\
\ub{Sr}CaSn &8.18 (8.42) &5.05 (5.17) &9.39 (9.69) &0.020 (0.021) &0.685 (0.681) &0.151 (0.151) &0.073 (0.071) &0.263 (0.262) &0.393 (0.390) &0.31\\
\ub{Sr}CaPb &8.22 (8.51) &5.08 (5.19) &9.43 (9.74) &0.020 (0.022) &0.685 (0.680) &0.152 (0.149) &0.073 (0.069) &0.263 (0.262) &0.393 (0.389) &0.31 \\
\ub{Ba}CaSi &8.37 (8.39) &4.93 (4.95) &8.92 (9.16) &0.020 (0.024) &0.682 (0.677) &0.145 (0.142) &0.074 (0.072) &0.276 (0.278) &0.392 (0.392) &0.04 \\
\ub{Ba}CaGe &8.37 (8.54) &4.94 (5.10) &8.97 (9.21) &0.027 (0.024) &0.675 (0.677) &0.141 (0.142) &0.075 (0.072) &0.279 (0.278) &0.389 (0.392) &0.05 \\
\ub{Ba}CaSn &8.66 (8.83) &5.18 (5.31) &9.44 (9.69) &0.020 (0.021) &0.682 (0.681) &0.145 (0.146) &0.074 (0.070) &0.276 (0.274) &0.392 (0.393) &0.14 \\
\ub{Ti}NiSi &6.08 (6.15) &3.61 (3.67) &6.95 (7.02) &0.024 (0.021) &0.681 (0.680) &0.141 (0.142) &0.060 (0.061) &0.265 (0.265) &0.377 (0.377) &$m$ \\
\ub{Ti}NiGe &6.16 (6.24) &3.69 (3.75) &7.04 (7.15) &0.025 (0.032) &0.687 (0.687) &0.139 (0.141) &0.060 (0.060) &0.257 (0.258) &0.380 (0.377) &$m$ \\
\ub{Ti}PdGe &6.26 (6.37) &3.79 (3.86) &7.39 (7.54) &0.034 (0.034) &0.685 (0.683) &0.142 (0.145) &0.060 (0.061) &0.244 (0.246) &0.376 (0.376) &$m$ \\
\ub{Ti}PtSi &6.27 (6.34) &3.75 (3.80) &7.25 (7.34) &0.027 (0.027) &0.675 (0.675) &0.147 (0.148) &0.063 (0.064) &0.258 (0.259) &0.376 (0.376) &$m$ \\
\ub{Zr}PtSi &6.54 (6.60) &3.86 (3.90) &7.47 (7.54) &0.024 (0.024) &0.678 (0.678) &0.144 (0.146) &0.063 (0.064) &0.265 (0.263) &0.378 (0.378) &$m$ \\
\ub{Zr}PtGe &6.59 (6.66) &3.93 (3.97) &7.57 (7.66) &0.026 (0.027) &0.682 (0.683) &0.143 (0.145) &0.063 (0.063) &0.257 (0.255) &0.380 (0.379) &$m$ \\
\ub{Hf}PdSi &6.43 (6.57) &3.79 (3.87) &7.39 (7.56) &0.027 (0.027) &0.680 (0.682) &0.142 (0.143) &0.061 (0.059) &0.260 (0.265) &0.377 (0.369) &$m$ \\
\ub{Hf}PtSi &6.44 (6.55) &3.82 (3.88) &7.39 (7.51) &0.025 (0.026) &0.678 (0.678) &0.145 (0.145) &0.064 (0.063) &0.260 (0.257) &0.377 (0.378) &$m$ \\
\ub{Hf}PtGe &6.49 (6.60) &3.90 (3.95) &7.50 (7.62) &0.028 (0.029) &0.683 (0.686) &0.144 (0.142) &0.064 (0.062) &0.252 (0.251) &0.379 (0.379) &$m$ \\
\end{tabular}
\caption{Computed structural parameters and band gap for $ABC$
  combinations experimentally determined to crystallize in the $Pnma$
  MgSrSi structure type.  All three elements occupy 4$c$ Wyckoff
  positions with free parameters $x$ and $z$.  The ``stuffing atom''
  $A$ is underlined. The last column is the computed band gap, where
  $m$ indicates that the combination is metallic. Experimental data
  are reported in parentheses.}
\label{table:knownortho}
\end{table}

\begin{table}
\begin{center}
\begin{tabular}{lcccccccccc}
$ABC$&$a$&$b$&$c$&$x_{\rm A}$&$z_{\rm A}$&$x_{\rm B}$&$z_{\rm B}$&$x_{\rm C}$&$z_{\rm C}$&$E_{\rm gap}$\\
     &(\AA)&(\AA)&(\AA)&&&&&&&(eV)\\
\ub{Na}AgSe &7.26 &4.23 &8.03 &0.035 &0.675 &0.121 &0.052& 0.253 &0.370 &0.11\\
\ub{Sr}LiAs &6.94 &5.68 &9.85 &0.005 &0.706 &0.156 &0.073& 0.279 &0.406 &1.00 \\
\ub{K}SrP   &8.12 &4.98 &8.68 &0.014 &0.675 &0.156 &0.087& 0.288 &0.405 &1.64 \\
\ub{K}SrSb  &6.94 &5.68 &9.75 &0.000 &0.750 &0.250 &0.083& 0.250 &0.417 &0.40  \\
\ub{K}SrBi  &7.04 &5.75 &9.97 &0.000 &0.750 &0.250 &0.083& 0.250 &0.417 &$m$ \\
\ub{Rb}SrAs &8.75 &5.17 &8.85 &0.013 &0.669 &0.150 &0.084& 0.298 &0.399 &1.39 \\
\ub{Rb}SrBi &9.17 &5.47 &9.60 &0.011 &0.675 &0.150 &0.082& 0.290 &0.398 &1.01 \\
\ub{Rb}BaAs &8.48 &5.42 &9.48 &0.009 &0.683 &0.164 &0.088& 0.283 &0.409 &1.16 \\
\ub{Sr}AuP  &8.32 &4.22 &7.33 &0.000 &0.750 &0.250 &0.084& 0.250 &0.418 &$m$  \\
\ub{La}LiSi &7.52 &4.47 &7.71 &0.002 &0.723 &0.188 &0.081& 0.275 &0.411 &$m$ \\
\ub{Na}InGe &7.75 &4.51 &8.03 &0.013 &0.672 &0.161 &0.083& 0.299 &0.394 &$m$  \\
\ub{Rb}InSn &8.92 &5.09 &8.27 &0.001 &0.665 &0.162 &0.091& 0.328 &0.402 &$m$ \\
\ub{Ba}SrSn &8.63 &5.33 &9.93 &0.023 &0.682 &0.149 &0.076& 0.262 &0.394 &0.07\\
\end{tabular}
\caption{Computed structural parameters and band gap for $ABC$
  combinations previously predicted to have a $Pnma$ ground state
  structure~\cite{Zhang12p1425}. Same description as preceding table.}
\label{table:zungerortho}
\end{center}
\end{table}

\begin{table}
\begin{center}
\begin{tabular}{lcccccccccc}
$ABC$&$a$&$b$&$c$&$x_{\rm A}$&$z_{\rm A}$&$x_{\rm B}$&$z_{\rm B}$&$x_{\rm C}$&$z_{\rm C}$&$E_{\rm gap}$\\
     &(\AA)&(\AA)&(\AA)&&&&&&&(eV)\\
\ub{Li}BeP  &6.09 &3.56 &6.38 &0.016 &0.676 &0.145 &0.070 &0.279 &0.390 &1.18\\
\ub{Mg}LiP  &6.44 &3.83 &7.37 &0.032 &0.680 &0.136 &0.063 &0.257 &0.381 &0.44\\
\ub{Li}ZnP  &6.41 &3.84 &7.02 &0.023 &0.667 &0.147 &0.070 &0.262 &0.389 &0.71\\
\ub{Li}BeAs &6.40 &3.74 &6.71 &0.016 &0.678 &0.145 &0.072 &0.276 &0.392 &1.35\\
\ub{Mg}LiAs &6.72 &3.98 &7.62 &0.034 &0.680 &0.135 &0.064 &0.257 &0.380 &0.35\\
\ub{Li}ZnAs &6.72 &4.00 &7.40 &0.026 &0.667 &0.141 &0.067 &0.257 &0.387 &0.80\\
\ub{Li}BeSb &6.89 &4.06 &7.19 &0.012 &0.690 &0.144 &0.078 &0.272 &0.400 &0.68\\
\ub{Mg}LiSb &7.18 &4.26 &8.19 &0.038 &0.680 &0.132 &0.063 &0.254 &0.378 &$m$\\
\ub{Li}ZnSb &7.16 &4.26 &7.82 &0.033 &0.678 &0.134 &0.063 &0.251 &0.386 &0.30\\
\ub{Li}BeBi &7.03 &4.13 &7.35 &0.015 &0.697 &0.138 &0.077 &0.267 &0.401 &0.39\\
\ub{Mg}LiBi &7.30 &4.35 &8.34 &0.038 &0.680 &0.132 &0.063 &0.254 &0.378 &$m$\\
\ub{Li}ZnBi &7.24 &4.31 &8.01 &0.049 &0.687 &0.123 &0.053 &0.238 &0.381 &$m$\\
\ub{Na}MgP  &7.30 &4.27 &7.73 &0.018 &0.667 &0.146 &0.074 &0.285 &0.389 &1.52\\
\ub{Na}MgAs &7.53 &4.40 &8.05 &0.019 &0.668 &0.146 &0.073 &0.282 &0.388 &0.97\\
\ub{Na}MgSb &7.89 &4.65 &8.58 &0.020 &0.670 &0.145 &0.070 &0.276 &0.386 &1.02\\
\ub{Na}ZnSb &7.73 &4.52 &7.87 &0.002 &0.683 &0.155 &0.084 &0.292 &0.401 &0.54\\
\ub{Na}MgBi &7.98 &4.71 &8.71 &0.020 &0.671 &0.143 &0.068 &0.273 &0.384 &0.11\\
\ub{Na}ZnBi &7.86 &4.62 &8.01 &0.001 &0.685 &0.155 &0.085 &0.292 &0.402 &$m$\\
\ub{K}MgSb  &8.44 &4.87 &8.48 &0.004 &0.672 &0.145 &0.074 &0.297 &0.391 &1.18\\
\ub{K}MgBi  &8.55 &4.95 &8.65 &0.005 &0.673 &0.144 &0.073 &0.295 &0.390 &0.30\\
\end{tabular}
\caption{Computed structural parameters and band gap for $ABC$
  combinations previously determined to be insulating and at least
  locally stable in the polar LiGaGe
  structure~\cite{Bennett12p167602}. Same description as
  preceding tables.}
\label{table:ourfeortho}
\end{center}
\end{table}

\begin{table}
\begin{center}
\begin{tabular}{lcccccccccc}
$ABC$    &  $a$& $c$& $E_{\rm gap}$& $a$& $c$& $z_{2a}$& $z_{2a\prime}$& $E_{\rm gap}$& $\Delta E_{\rm SW}$& $\Delta E$\\
         & (\AA)& (\AA)& (eV)&(\AA)&(\AA)&&&(eV)&(eV)&(eV)\\
\ub{Na}LiSe &4.25 &7.37 &2.66 &4.25 &7.37 &0.250 &0.750 &2.66 &0 &0\\
\ub{Na}LiTe &4.65 &7.66 &2.71 &4.67 &7.34 &0.308 &0.724 &2.79 &0.022 &0.113\\  
\ub{K}NaS   &4.72 &7.24 &2.30 &4.70 &7.27 &0.288 &0.720 &2.49 &0.012 &0.133\\
\ub{K}NaSe  &4.93 &7.38 &1.89 &4.92 &7.45 &0.290 &0.722 &2.03 &0.013 &0.131\\ 
\ub{K}NaTe  &5.29 &7.61 &2.03 &5.25 &7.87 &0.296 &0.725 &2.21 &0.014 &0.096\\ 
\ub{Ca}LiSb &4.59 &7.52 &0.21 &4.60 &7.44 &0.282 &0.733 &0.27 &0.004 &0.079\\ 
\ub{Ca}LiBi &4.67 &7.57 &$m$ &4.68 &7.51 &0.286 &0.731 &0.15 &0.007 &0.073\\
\ub{Sr}LiSb &4.69 &8.19 &0.62 &4.69 &8.19 &0.250 &0.750 &0.62 &0 &0\\ 
\ub{Sr}LiBi &4.77 &8.23 &0.45 &4.77 &8.23 &0.250 &0.750 &0.45 &0 &0\\  
\ub{Mg}CuP  &3.87 &7.17 &$m$ &3.96 &6.52 &0.300 &0.700 &0.36 &0.001 &0.175\\
\ub{Na}CdAs &4.51 &7.88 &$m$ &4.56 &7.22 &0.295 &0.686 &0.26 &0.199 &0.050\\
\ub{Na}CdSb &4.86 &7.50 &$m$ &4.79 &7.58 &0.312 &0.701 &0.24 &0.282 &-0.037\\
\ub{Sc}NiP  &3.76 &6.98 &$m$ &3.76 &6.98 &0.258 &0.743 &$m$ &0.001 &0.268\\
\ub{Sc}PtP  &3.95 &7.16 &$m$ &3.96 &7.09 &0.264 &0.726 &$m$ &0.008 &0.254\\
\ub{Sc}CuSi &4.03 &6.53 &$m$ &4.05 &6.40 &0.302 &0.719 &$m$ &0.124 &0.003\\
\ub{Ca}MgSi &4.63 &6.99 &$m$ &4.62 &7.09 &0.277 &0.711 &$m$ &0.025 &0.197\\
\ub{Ca}MgGe &4.67 &6.95 &$m$ &4.65 &7.11 &0.278 &0.710 &$m$ &0.028 &0.193\\ 
\ub{Ca}MgSn &4.93 &7.17 &$m$ &4.91 &7.34 &0.284 &0.716 &$m$ &0.027 &0.119\\ 
\ub{Sr}MgSi &4.69 &7.75 &$m$ &4.71 &7.71 &0.273 &0.701 &$m$ &0.015 &0.195\\
\ub{Sr}MgGe &4.74 &7.66 &$m$ &4.75 &7.66 &0.276 &0.698 &$m$ &0.027 &0.199\\
\ub{Sr}MgSn &5.00 &7.81 &$m$ &5.00 &7.86 &0.286 &0.708 &$m$ &0.034 &0.119\\
\ub{Sr}CaSi &5.07 &7.28 &$m$ &5.07 &7.28 &0.254 &0.744 &$m$ &0.180 &0.218\\
\ub{Sr}CaGe &5.09 &7.27 &$m$ &5.09 &7.24 &0.253 &0.745 &$m$ &0.002 &0.185\\
\ub{Sr}CaSn &5.36 &7.39 &$m$ &5.36 &7.86 &0.250 &0.750 &$m$ &0 &0\\
\ub{Sr}CaPb &5.39 &7.45 &$m$ &5.39 &7.45 &0.250 &0.750 &$m$ &0 &0\\
\ub{Ba}CaSi &5.16 &7.90 &$m$ &5.13 &8.05 &0.272 &0.707 &0.05 &0.011 &0.310\\
\ub{Ba}CaGe &5.20 &7.78 &$m$ &5.17 &7.94 &0.270 &0.717 &$m$ &0.010 &0.289\\
\ub{Ba}CaSn &5.46 &7.95 &$m$ &5.44 &8.06 &0.267 &0.726 &$m$ &0.002 &0.227\\
\ub{Ti}NiSi &4.03 &5.37 &$m$ &3.90 &5.93 &0.325 &0.733 &$m$ &0.236 &0.161\\
\ub{Ti}NiGe &4.10 &5.40 &$m$ &3.97 &5.99 &0.330 &0.736 &$m$ &0.221 &0.008\\ 
\ub{Ti}PdGe &4.27 &5.48 &$m$ &4.14 &6.06 &0.315 &0.729 &$m$ &0.036 &0.242\\
\ub{Ti}PtSi &4.21 &5.47 &$m$ &4.09 &6.07 &0.318 &0.727 &$m$ &0.166 &0.281\\
\ub{Zr}PtSi &4.30 &5.93 &$m$ &4.18 &6.49 &0.316 &0.727 &$m$ &0.345 &0.390\\ 
\ub{Zr}PtGe &4.38 &5.88 &$m$ &4.25 &6.51 &0.324 &0.729 &$m$ &0.357 &0.179\\ 
\ub{Hf}PdSi &4.11 &6.33 &$m$ &4.12 &6.37 &0.316 &0.726 &$m$ &0.314 &0.378\\
\ub{Hf}PtSi &4.10 &6.46 &$m$ &4.14 &6.41 &0.319 &0.727 &$m$ &0.337 &0.274\\
\ub{Hf}PtGe &4.23 &6.20 &$m$ &4.21 &6.43 &0.327 &0.730 &$m$ &0.431 &0.050\\  
\end{tabular}
\caption{Computed structural parameters, total energy and band gap for
  nonpolar $P6_{3}/mmc$ (194) and polar $P6_{3}mc$ (186) phases for
  $ABC$ combinations experimentally determined to crystallize in the
  $Pnma$ MgSrSi structure type.  $ABC$ combinations, in which the
  experimental ground state is reported to be MgSrSi structure type,
  in hexagonal symmetries to screen for stable ferroelectric
  distortions. The lattice constants and band gap for nonpolar
  $P6_{3}/mmc$ (194) symmetry are given in the first three
  columns. The lattice constants, Wyckoff parameters, and band gap for
  polar $P6_{3}mc$ (186) symmetry are given in the next five
  columns. The ferroelectric switching barrier ($\Delta E_{\rm SW}$,
  from $P6_{3}/mmc$ to $P6_{3}mc$) and the difference in energy
  ($\Delta E$) between $Pnma$ and $P6_{3}mc$ are given in the last two
  columns. Note that if $\Delta E_{\rm SW}$ is zero, then polar
  $P6_{3}mc$ symmetry is not stable, and thus $\Delta E$ is also
  zero.}
\label{table:knownashex}
\end{center}
\end{table}

\begin{table}
\begin{center}
\begin{tabular}{lcccccccccc}
$ABC$    &  $a$& $c$& $E_{\rm gap}$& $a$& $c$& $z_{2a}$& $z_{2a\prime}$& $E_{\rm gap}$& $\Delta E_{\rm SW}$& $\Delta E$\\
         & (\AA)& (\AA)& (eV)&(\AA)&(\AA)&&&(eV)&(eV)&(eV)\\
\ub{Na}AgSe &4.47 &7.44 &$m$ &4.48 &7.08 &0.297 &0.698 &0.05 &0.051 &0.133\\ 
\ub{Sr}LiAs &4.41 &7.95 &1.08 &4.41 &7.95 &0.250 &0.750 &1.07 &0 &0\\ 
\ub{K}SrP   &5.25 &6.47 &0.67 &5.25 &6.47 &0.250 &0.750 &0.67 &0 &0\\ 
\ub{K}SrSb  &5.69 &6.94 &0.40 &5.69 &6.94 &0.250 &0.750 &0.40 &0 &0\\ 
\ub{K}SrBi  &5.76 &7.04 &$m$ &5.76 &7.04 &0.250 &0.750 &$m$ &0 &0\\ 
\ub{Rb}SrAs &5.49 &6.73 &0.79 &5.49 &6.73 &0.250 &0.750 &0.79 &0 &0\\ 
\ub{Rb}SrBi &5.86 &7.12 &0.16 &5.86 &7.12 &0.250 &0.750 &0.17 &0 &0\\ 
\ub{Rb}BaAs &5.67 &6.98 &0.49 &5.67 &6.98 &0.250 &0.750 &0.49 &0 &0\\
\ub{Sr}AuP  &4.18 &8.51 &$m$ &4.18 &8.51 &0.250 &0.750 &$m$ &0 &0\\ 
\ub{La}LiSi &4.46 &7.66 &$m$ &4.46 &7.64 &0.263 &0.733 &$m$ &0.002 &0.022\\ 
\ub{Na}InGe &4.98 &6.03 &$m$ &4.64 &7.23 &0.291 &0.683 &$m$ &0.312 &0.001\\ 
\ub{Rb}InSn &5.22 &7.91 &$m$ &5.20 &7.86 &0.301 &0.699 &$m$ &0.333 &0.267\\
\ub{Ba}SrSn &5.68 &7.75 &$m$ &5.68 &7.76 &0.250 &0.750 &$m$ &0 &0\\ 
\end{tabular}
\caption{Computed structural parameters, total energy and band gap for
  nonpolar $P6_{3}/mmc$ (194) and polar $P6_{3}mc$ (186) phases for
  $ABC$ combinations previously predicted to have a $Pnma$ ground
  state structure~\cite{Zhang12p1425}.  Same description as preceding
  table. Note that if $\Delta E_{\rm SW}$ is zero, then polar
  $P6_{3}mc$ symmetry is not stable, and thus $\Delta E$ is also
  zero. }
\label{table:knownashex}
\end{center}
\end{table}

\begin{table}
\begin{center}
\begin{tabular}{lcccccccccc}
$ABC$    &  $a$& $c$& $E_{\rm gap}$& $a$& $c$& $z_{2a}$& $z_{2a\prime}$& $E_{\rm gap}$& $\Delta E_{\rm SW}$& $\Delta E$\\
         & (\AA)& (\AA)& (eV)&(\AA)&(\AA)&&&(eV)&(eV)&(eV)\\
\ub{Li}BeP  &3.56 &6.99 &1.55 &3.63 &5.83 &0.295 &0.686 &1.51 &0.119 &0.018\\
\ub{Mg}LiP  &4.03 &6.41 &0.73 &4.03 &6.42 &0.303 &0.711 &0.60 &0.019 &0.230\\
\ub{Li}ZnP  &3.94 &6.78 &1.42 &3.92 &6.36 &0.341 &0.722 &1.33 &0.323 &0.059\\
\ub{Li}BeAs &3.75 &7.26 &1.66 &4.09 &6.64 &0.265 &0.647 &1.71 &0.148 &0\\
\ub{Mg}LiAs &4.20 &6.69 &0.53 &4.19 &6.68 &0.302 &0.712 &0.48 &0.029 &0.207\\
\ub{Li}ZnAs &4.16 &6.79 &0.42 &4.11 &6.67 &0.335 &0.715 &0.97 &0.360 &0\\
\ub{Li}BeSb &4.06 &7.64 &1.08 &4.09 &6.65 &0.267 &0.650 &1.13 &0.288 &0\\ 
\ub{Mg}LiSb &4.53 &6.82 &$m$ &4.50 &6.98 &0.301 &0.715 &$m$ &0.050 &0.142\\
\ub{Li}ZnSb &4.44 &7.19 &$m$ &4.38 &7.08 &0.288 &0.669 &0.67 &0.400 &0\\
\ub{Li}BeBi &4.14 &7.78 &1.02 &4.18 &6.81 &0.262 &0.643 &0.45 &0.311 &0\\
\ub{Mg}LiBi &4.64 &6.88 &$m$ &4.60 &7.05 &0.300 &0.718 &$m$ &0.041 &0.126\\
\ub{Li}ZnBi &4.76 &6.18 &$m$ &4.46 &7.23 &0.277 &0.658 &$m$ &0.309 &0\\  
\ub{Na}MgP  &4.41 &7.09 &1.65 &4.42 &6.88 &0.310 &0.716 &1.17 &0.102 &0.275\\
\ub{Na}MgAs &4.58 &7.09 &0.77 &4.55 &7.26 &0.314 &0.715 &0.67 &0.114 &0.232\\
\ub{Na}MgSb &4.90 &7.46 &0.09 &4.87 &7.58 &0.307 &0.705 &0.69 &0.146 &0.154\\
\ub{Na}ZnSb &4.47 &8.76 &0.82 &4.56 &7.33 &0.285 &0.677 &0.20 &0.081 &0.042\\ 
\ub{Na}MgBi &4.99 &7.40 &$m$ &4.96 &7.58 &0.302 &0.704 &0.14 &0.127 &0.143\\
\ub{Na}ZnBi &4.92 &6.90 &$m$ &4.65 &7.48 &0.283 &0.675 &$m$ &0.328 &0.018\\ 
\ub{K}MgSb  &4.86 &9.25 &1.02 &5.02 &7.79 &0.290 &0.697 &0.59 &0.041 &0.254\\  
\ub{K}MgBi  &4.95 &9.30 &0.34 &5.09 &8.01 &0.291 &0.698 &0.15 &0.073 &0.227\\ 
\end{tabular}
\caption{Computed structural parameters, total energy and band gap for
  nonpolar $P6_{3}/mmc$ (194) and polar $P6_{3}mc$ (186) phases for
  $ABC$ combinations previously determined to be insulating and at
  least locally stable in the polar LiGaGe
  structure. ~\cite{Bennett12p167602}. Same description as
  preceding tables. If $\Delta E$ is zero, the polar $P6_{3}mc$
  structure is predicted to be the ground state, and not the antipolar
  $Pnma$ structure.}
\label{table:ourknownashex}
\end{center}
\end{table}

\begin{table}
\begin{center}
\begin{tabular}{lcccccccccc}
$ABC$    &  $Space Group$ & $a^{\rm AE}$& $b^{\rm AE}$& $c^{\rm AE}$& $a^{\rm PP}$&  $b^{\rm PP}$&  $c^{\rm PP}$&  $\Delta a$ &  $\Delta b$&  $\Delta c$ \\
         &  &(\AA)& (\AA)& (\AA)&(\AA)& (\AA)& (\AA)& (\%)& (\%)& (\%)\\
 \ub{Ca}LiBi  & $Pnma$ & 7.58 & 4.63 & 8.23 & 7.57 & 4.59 & 8.21    & -0.15 & -0.76 & -0.18 \\
  \ub{Sr}LiBi & $Pnma$ & 7.98 & 4.81 & 8.40 & 7.95 & 4.78 & 8.37    & -0.35 & -0.62 & -0.36 \\
  \ub{Ca}LiSb & $Pnma$ & 7.45 & 4.54 & 8.06 & 7.46 & 4.53 & 8.07    & 0.13 &  -0.22 & 0.12  \\
  \ub{Sc}PtP  & $Pnma$ & 6.45 & 3.81 & 7.35 & 6.47 & 3.78 & 7.30    & 0.25 &  -0.87 & -0.67 \\
  \ub{K}NaTe  & $Pnma$ & 8.27 & 4.96 & 9.10 & 8.35 & 5.00 & 9.26    & 0.91 &  0.83  & 1.71  \\
  \ub{Na}CdAs & $Pnma$ & 7.39 & 4.37 & 7.90 & 7.49 & 4.42 & 8.03    & 1.37 &  1.07  & 1.70 \\ 
  \ub{Ca}LiBi & $P6_{3}/mmc$ & 4.67 & - & 7.57 &  4.71 & - &7.50   & -0.87 & - & 0.91 \\
  \ub{Sr}LiBi & $P6_{3}/mmc$ & 4.77 & - & 8.23 &  4.79 & - &8.25   & -0.47 & - &-0.19 \\
  \ub{Ca}LiSb & $P6_{3}/mmc$ & 4.60 & - & 7.52 &  4.60 & - &7.46   & -0.18 & - & 0.91 \\
  \ub{Sr}CaSi & $P6_{3}/mmc$ & 5.07 & - & 7.28 &  5.07 & - &7.26   &  0.00 & - & 0.20 \\
\end{tabular}
\caption{All-electron Wien2K~\cite{Schwarz03p259} (AE) lattice
constants for selected structures are compared to the pseudopotential
(PP) results described in the manuscript. Wien2K $Pnma$ lattice
constants were calculated with internal parameters fixed to results
from PP calculations.}
\label{table:w2k}
\end{center}
\end{table}

\bibliography{thebibliography}